\documentclass[aps,prl,amsmath,amssymb,reprint,superscriptaddress]{revtex4-1}
\usepackage{graphicx}
\usepackage{amsmath}
\usepackage{dcolumn}
\usepackage{bm}
\usepackage{bbold}
\usepackage{color}
\usepackage{amsfonts}
\usepackage{amssymb}
\usepackage{mathrsfs}
\usepackage{tabularx}
\usepackage{braket}
\usepackage{mathtools}
\usepackage{soul}			

\usepackage{caption}
\usepackage{subcaption}

\begin{document}

\title{Indirect Methods to Control Population Distribution in a Large Spin System}

\author{Lingfei Zhao}
\affiliation{Department of Physics, Nanjing University, Nanjing 210093, People's Republic of China}
\author{Maxim Goryachev}\affiliation{ARC Centre of Excellence for Engineered Quantum Systems, School of Physics, University of Western Australia, 35 Stirling Highway, Crawley WA 6009, Australia}
\author{Jeremy Bourhill}\affiliation{ARC Centre of Excellence for Engineered Quantum Systems, School of Physics, University of Western Australia, 35 Stirling Highway, Crawley WA 6009, Australia}
\author{Michael E. Tobar}\affiliation{ARC Centre of Excellence for Engineered Quantum Systems, School of Physics, University of Western Australia, 35 Stirling Highway, Crawley WA 6009, Australia}

\date{\today}

\begin{abstract}
We demonstrate how a large spin system ($S=7/2$) with the ground and the first excited state separated by a seven photon transition exhibit non equilibrium thermodynamic properties and how the population distribution may be manipulated using coupling between energy levels. The first method involves non-adiabatic passage through an avoided level crossing controlled with an external DC magnetic field and the resulting Landau Zener transition. The second method is based on external cavity pumping to a higher energy state hybridised with another state that is two single photon transitions away from the ground state. The results are confirmed experimentally with Gd$^{3+}$ impurity ion ensemble in a YVO$_{4}$ crystal cooled to 20 mK, which also acts as a microwave photonic whispering gallery mode resonator. Extremely long life times are observed due to large number of photons required for the transition between the ground and the first excited states.
 
\end{abstract}

\maketitle


\section{Introduction}

{Physical systems based on the light-matter interaction, in general, and quantum electrodynamics (QED) with {``}spins-in-solid{''}, in particular, have found very broad range of applications. They constitute a platform for lasers\cite{lasersss} and masers\cite{maser, Creedon:2010aa}, quantum information processing units\cite{Pla:2012aa,BertainaS.:2007aa} and quantum communication devices\cite{OBrien:2014aa,Williamson:2014aa}, clocks\cite{Nicholson:2015aa,Rogge:2013aa} and sensors\cite{Maze:2008aa}, as well as probes in fundamental physics\cite{Bluhm:2000aa,Van-Tilburg:2015aa}. Particular interest is devoted to Rare-Earth spin ensembles in crystals due to their long lifetimes\cite{Probst:2013aa,Tkalcec:2014aa,PhysRevB.78.085410}. Although many aspects of such systems have been investigated in detail, many other features are yet to be discovered. Such new peculiarities not only give birth to new generations of existing devices and tools but also open avenues for new applications. 

In this work, we experimentally demonstrate a new configuration spin system that can bring new ideas to the field of quantum electrodynamics with spins. In particular, we consider an unusual manifestation of the spin angular momentum conservation law appearing through unusual population dynamics in a multilevel spin system. Although it is well understood that light-matter interaction for small enough coupling (not reaching the ultra-strong coupling regime, i.e. when coupling is small enough comparing to the resonance frequencies involved) is a subject of conservation of energy and momentum, the role of the latter is not usually discussed though it may play a very crucial role in many systems. The spin angular momentum conservation law imposes an additional selection rule for spin-photon interaction by creating photons with circular polarisation and thus the spin angular momentum. For instance, it shapes the system response for a photon interacting with a spin ensemble in a Whispering Gallery Mode (WGM) cavity by breaking the time reversal symmetry\cite{Goryachev:2014aa,Goryachev:2014ab}. The quantum system configuration in such system may be significantly different to that of traditional QED systems giving new properties that can be exploited for existing and emerging applications.
}


A typical setup implementing spin-photon interaction consists of a cavity coupled to a two-level system (TLS) or an ensemble of TLS\cite{Verdu:2009aa,Amsuss:2011aa,Lukin:2003aa,maser, PhysRevApplied.6.024021}. Each TLS may absorb or emit one photon by transiting between the excited and the ground states. In this case, these states are separated by spin number one ($\Delta m = 1$). On the other hand, the ground and the first excited states may be separated by more than one spin number ($\Delta m > 1$). In this case, the transition requires the corresponding number of photons to conserve the spin number making  the transition less probable as $\Delta m$ photons are simultaneously required to interact with a spin. This fact not only reduces spin-cavity coupling but also spin interaction with the environment. As a result, spins excited to the first excited states exhibit significantly lower decay rate. On the other hand, despite lower coupling to the cavity, the spin state can be manipulated via coupling to other excited states. This work demonstrates the possibility of manipulation of such system via cavity pumping to higher excited states and the Landau Zener effect in a {``}spins-in-solid{''} experimental setup.

\section{Gd$^{3+}$: YVO$_{4}$ Spin System}

\begin{figure}[htbp]
\centering
\includegraphics[width=0.8\columnwidth]{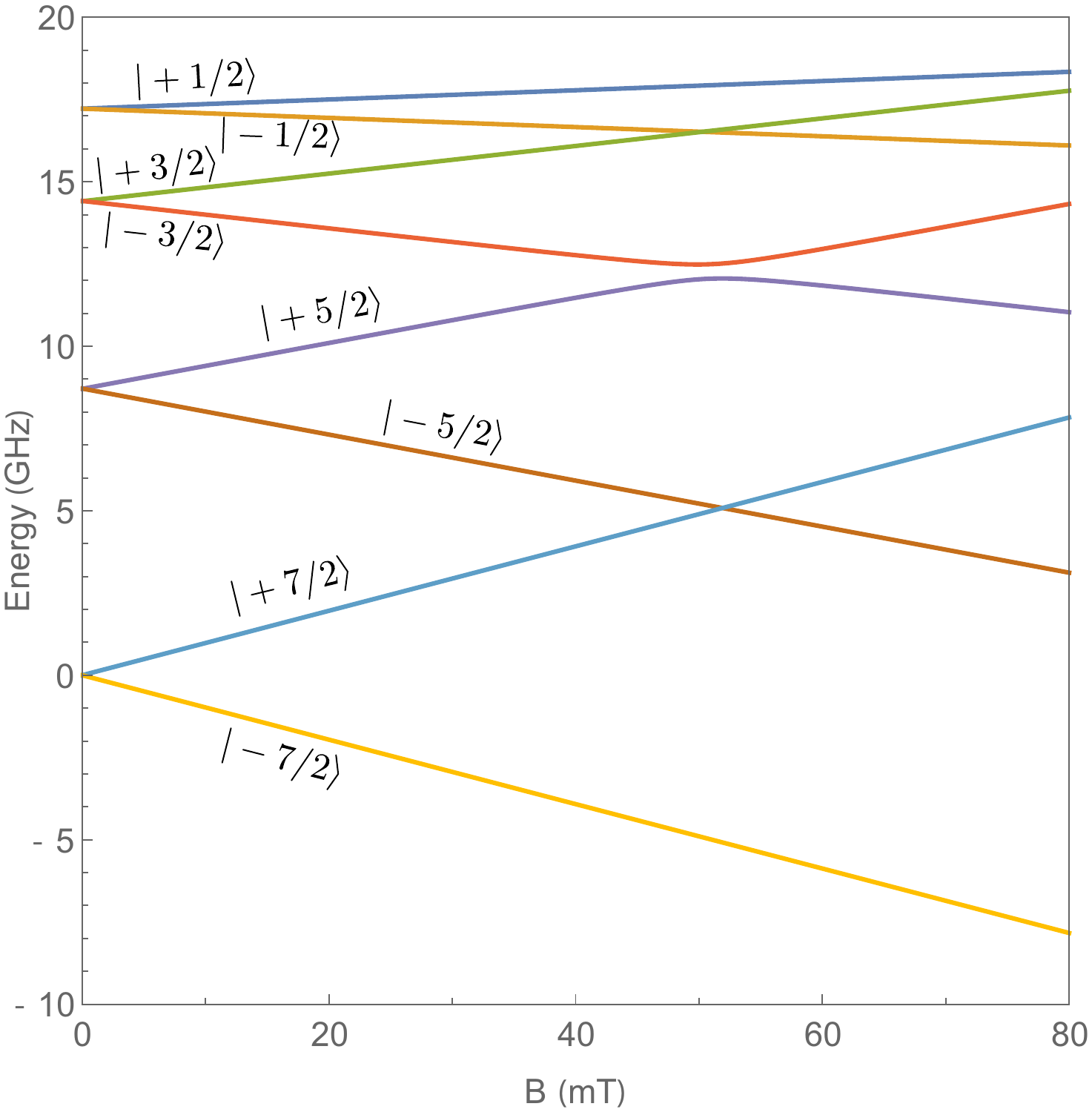}
\caption{Energy levels of Gd$^{3+}$ in YVO$_{4}$ as a function of magnetic field $B$ parallel to the crystal c-axis.}
\label{energylevel}
\end{figure}

The multi-photon transition spin system is realised with the Gd$^{3+}$ impurity ion ensemble in YVO$_{4}$ crystal having the zircon structure\cite{Wang:2004aa}. This impurity is a $7/2$ spin system with 8 energy levels controllable via an external magnetic field via Zeeman splitting\cite{Rosenthal:1969aa}. At zero field, the energy structure consists of four degenerate pairs of levels in the microwave frequency range (Fig.~\ref{energylevel})\cite{Urban:1968aa,Kahle:1968aa}. The most important feature of such a system is that the two lowest energy levels are $\ket{-7/2}$ and $\ket{+7/2}$ spin states ($\Delta m = 7$) requiring 7 photons for a transition. It is this large spin number difference that is exploited in this work to lock the spin system in the first excited state ($\ket{+7/2}$ state). The $\ket{-7/2}$ state plays the role of the ground state throughout this work. 

The other important feature of the Gd$^{3+}$:YVO$_{4}$ spin system is finite coupling between different energy levels. The origins of this coupling can vary from imperfections of the crystal structure to imperfect alignment of the external field with respect to the c-axis. This coupling is exploited for spin transfer between states with different spin number with respect to the ground state. 

In the following theoretical and experimental sections, we limit the discussion to the case of the lowest temperature routinely achievable with dilution refrigerators $T = 20$mK. This temperature is low enough to keep the whole spin ensemble at the lowest energy state in thermal equilibrium.


\subsection{Preparing a Non Equilibrium State}

\begin{figure}[htbp]
\centering
\includegraphics[width=0.58\columnwidth]{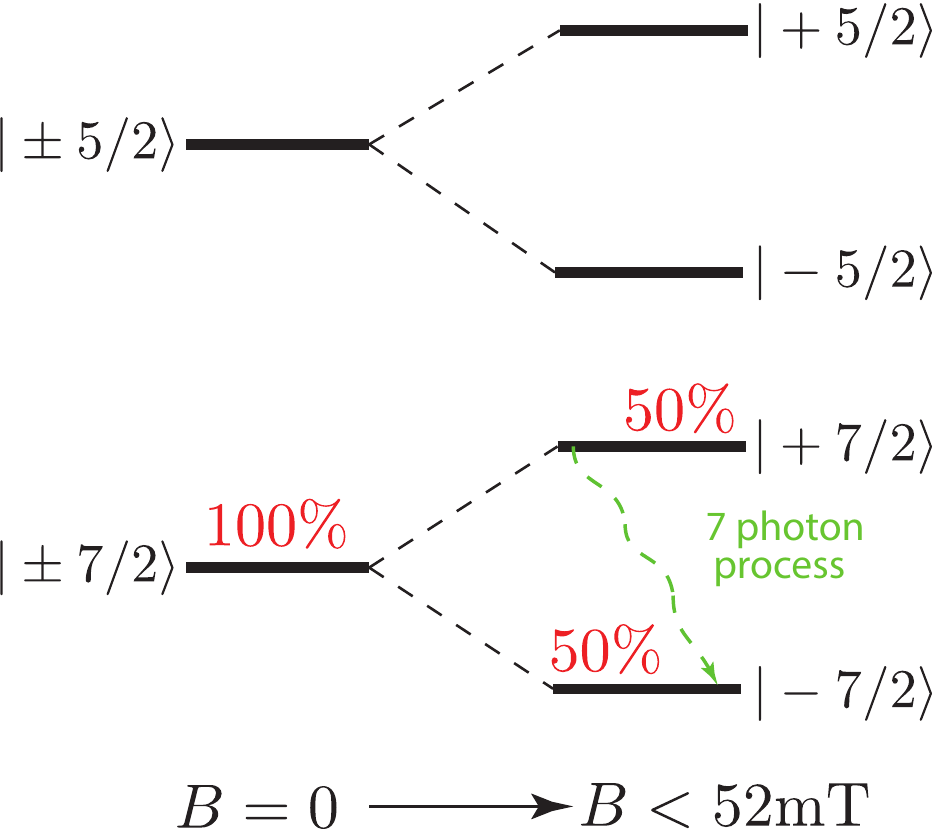}
\caption{Generation of the equally distributed state and the 7 photon process. Only $\ket{\pm7/2}$ and $\ket{\pm5/2}$ are shown. Percentages show the corresponding level occupancy.}
\label{figA}
\end{figure}

At zero magnetic field and for temperatures approaching absolute zero, population of the Gd$^{3+}$ spin ensemble condenses to the degenerate $\ket{\pm 7/2}$ states at the thermal equilibrium. This means that the spin ensemble has an equal distribution between these states. For nonzero magnetic fields, the thermal equilibrium is deviated from this 50-50\% distribution approaching complete 100\% occupation of the ground state. For a normal $\Delta m=1$ spin system, this thermalisation would happen naturally on the time scale corresponding to the strength of coupling to the environment. For the seven photon transition however, half of the population is naturally locked at the $\ket{+7/2}$ state and cannot be easily thermalised. Thus, the overall system exists in a non-thermal equilibrium state with equal distribution as the external magnetic field is initially slowly detuned from zero. This situation is bounded from the top at $B=52$mT where energy level $\ket{+7/2}$ and ${\ket{-5/2}}$ come to close proximity with one another. The overall process is shown in Fig.~\ref{figA}.

Occupancy of $\ket{\pm 7/2}$ states may be verified by measuring the coupling strength between a cavity mode and spin transitions from the corresponding states to some higher energy states. Indeed, the effective coupling strength for a $\ket{n/2}$ state is $g_\textrm{eff}= g_{0} \sqrt{N_{\ket{n/2}}}$, where $g_{0}$ is the coupling strength of an individual spin and $N_{\ket{n/2}}$ is the number of spins in the corresponding state under the given experimental conditions. $g_{0}$ is determined by spin properties together with the other parameters such as cavity mode filling factors\cite{goryachev2015discovery,Probst:2014aa}:
\begin{equation} 
g_{0}=\frac{g_{d} \, \mu_{B}}{2h} \sqrt{\frac{h f_{0} \mu \beta}{2V}},
\label{g0}
  \end{equation}
where $g_{d}$ is the Land\'{e} g-factor, $\mu_{B}$ is Bohr magneton, $h$ is Planck constant, $f_{0}$ is the frequency of the cavity mode, $\mu$ is the permeability of the crystal, $\beta$ is the magnetic field filling factor of the cavity mode which can be calculated by finite element analysis, and $V$ is the volume of the crystal.

\subsection{Landau-Zener Assisted Thermalisation}

The energy diagram of a Gd$^{3+}$ ion in YVO$_4$ suggests that for the external magnetic field near $52$mT energy levels $\ket{+7/2}$ and $\ket{-5/2}$ cross one another. This picture holds true only for an ideal crystal with external field exactly parallel to the crystal c-axis. In a real experiment this is not the case, and small misalignments between these directions cause coupling $g_{LZ}$ between $\ket{+7/2}$ and $\ket{-5/2}$ energy levels. As a result of this coupling, the levels never cross and form an Avoided Level Crossing (ALC), with the whole system  tuneable via the external $B$-field.

\begin{figure}[htbp]
\centering
\includegraphics[width=0.58\columnwidth]{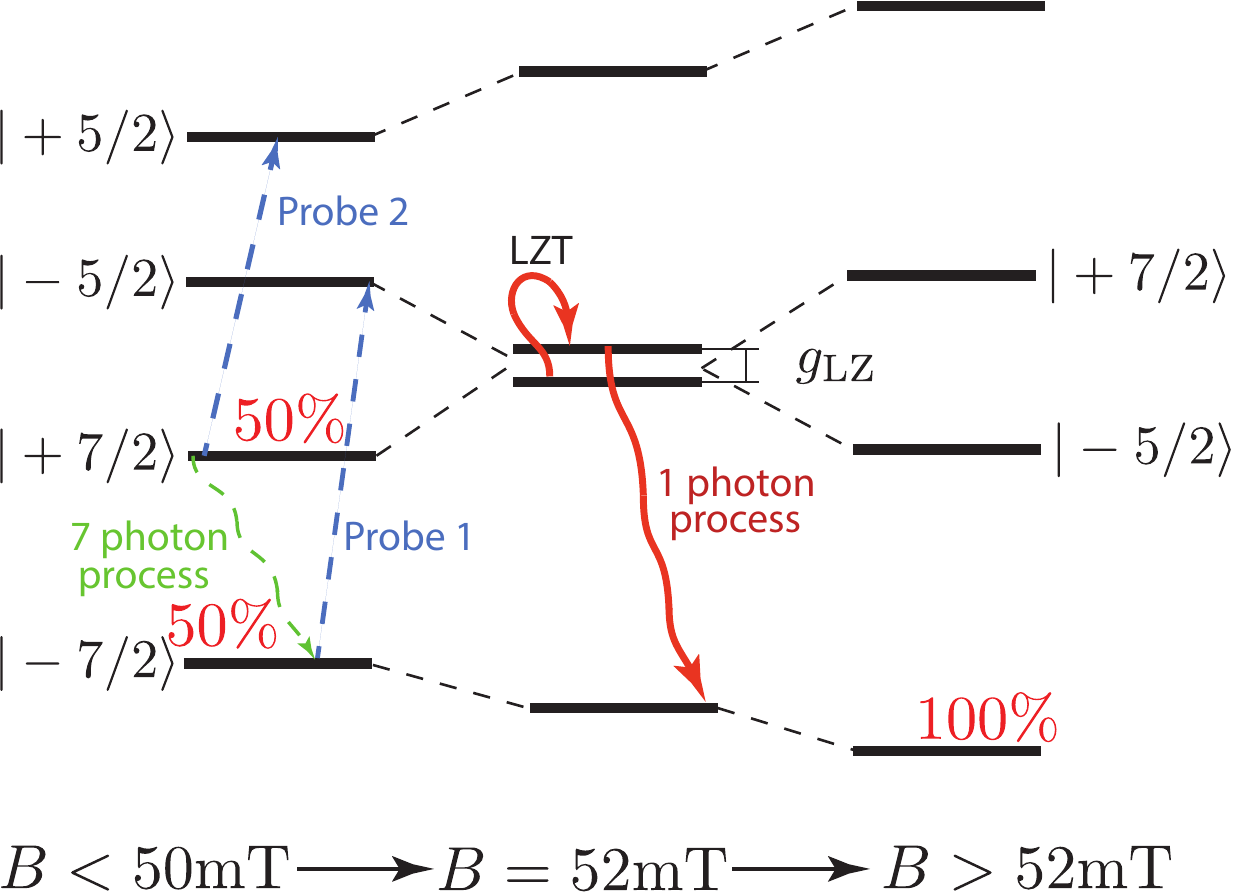}
\caption{Ensemble thermalisation through the Landau-Zener effect. Only four lower energy levels are shown.}
\label{figB}
\end{figure}

In the situation when energy levels are coupled and the external control parameter is tuned across the ALC, a quantum system exhibits quantum tunnelling Landau-Zener (LZ) transition\cite{Landau,Zener:1932aa}. In the discussed system, the tunnelling occurs from the populated $\ket{+7/2}$ state to the empty $\ket{-5/2}$. The latter is only a single photon transition away from the lowest energy state $\ket{-7/2}$. As a result,  the relaxation to this state occurs on the time scale of the system relaxation time. Consequently, the whole population of $\ket{-5/2}$, i.e. 50\% of the total ensemble (see right situation in Fig.~\ref{figB}), relax back to the ground $\ket{-7/2}$ state forming the thermal equilibrium state via the LZ transition. This process occurs when the control parameter is tuned across $52$mT (middle situation in Fig.~\ref{figB}). Thus, the 7 photon population, which has 50\% of the spins locked at the first excited stated is overcome by the non-adiabatic crossing of energy levels, and a normal one photon relaxation occurs requiring no external pumping field.

The Landau-Zener assisted thermalisation may be detected by measuring the coupling between a cavity mode and the spin transitions from both the $\ket{+7/2}$ and $\ket{-7/2}$ levels to $\ket{+5/2}$ and $\ket{-5/2}$, respectively. Before the event, the population will be equally split between the two lowest levels, and afterwards, all ions will be in the ground state as depicted in Fig.~\ref{figB}. Real time observation of this effect is not possible in our cavity implementation as there does not exist a cavity mode of frequency equivalent to the transition energy at $B=52$ mT.

\subsection{Cavity Driven Thermalisation}

\begin{figure}[htb]
\centering
\includegraphics[width=0.58\columnwidth]{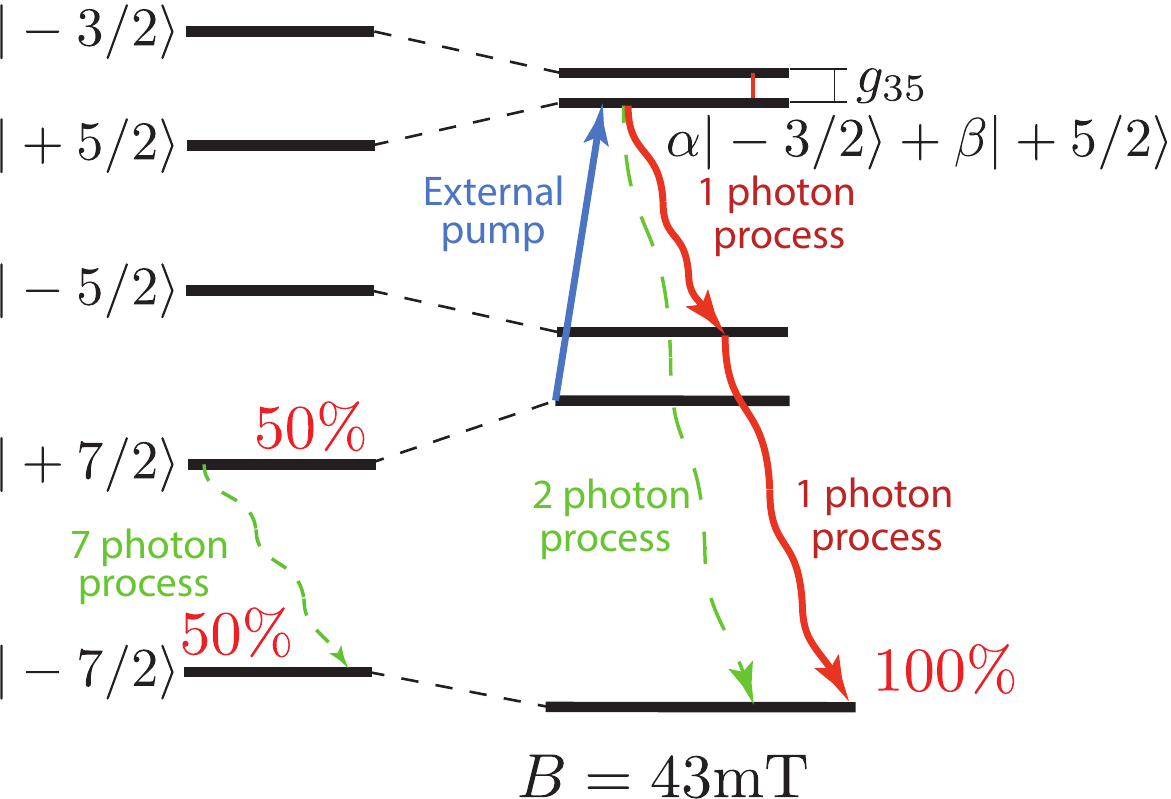}
\caption{Cavity assisted thermalisation via pumping to hybridised $\alpha \ket{+5/2} + \beta \ket{-3/2}$ state and decay to the ground state through $\ket{-5/2}$.}
\label{FigC}
\end{figure}

In addition to the Landau-Zener mechanism, the population can be transferred via an external pump field. In fact, one can pump the population at the $\ket{+7/2}$ state in Fig.~\ref{figA} (final situation) to a higher energy state that is closer to the ground state in number of photons required for relaxation. For the case of Gd$^{3+}$ in YVO$_4$, the procedure may be implemented as follows: the population from the $\ket{+7/2}$ state is pumped to the $\ket{+5/2}$ (1 photon efficient process), which is hybridised with $\ket{-3/2}$ at $B = 43$~mT, ions at the $\ket{-3/2}$ state decay to $\ket{-7/2}$ via $\ket{-5/2}$ (two single photon processes). The energy level diagram for the cavity driven thermalisation is shown in Fig.~\ref{FigC}.
The coupling between $\ket{+5/2}$ and $\ket{-3/2}$ states arises due to the fourthfold symmetry of the crystal and may exceed $100$~MHz\cite{Urban:1968aa}. At the working $43$ mT, the hybridised state is $0.05 \ket{-3/2} + 0.95 \ket{+5/2}$, which together with the splitting of the decay into two single photon processes may decrease the efficiency of the thermalisation. On the other hand, the overall process efficiency may be controlled via the strength of the external pump. Indeed, the time required to excite all the spins from $\ket{+7/2}$ to the hybridised higher energy state depends on the spin-photon coupling and the number of cavity photons. 

Cavity driven thermalisation may be observed in real time as it does not require any tuning parameters to be involved.  For this purpose, one can monitor the cavity transmission when it is tuned on the $\ket{+7/2}$ to $\alpha \ket{-3/2} + \beta \ket{+5/2}$ at $B=42$~mT as it is a function of $g_\textrm{eff}$.

\section{Experimental Realization}

To observe the effects described in the previous section, the {``}spins-in-solids{''} approach\cite{Kubo:2010aa,Schuster:2010aa,Amsuss:2011aa} with a WGM system\cite{Farr:2013aa} is implemented. The system utilises a single cylindrical YVO$_4$ crystal both as a cavity and ion host. This ensures the highest filling and quality factors.
The YVO$_{4}$ crystal has a diameter of 13.9~mm and height of 14.3~mm that provides several WGMs in the 5-20 GHz frequency range. The spin-photon interaction is maximised for WGMs with radial component of the magnetic field (WGH modes).
The crystal is confined inside an oxygen-free copper cavity and cooled to $20$~mK inside a dry dilution refrigerator  system. A superconducting magnet is used to provide a DC magnetic field almost parallel to the crystal c-axis. The external pump and probe fields are provided by a network analyzer connected to the cavity via electric field probes through a chain of cold attenuation (40dB in total). The transmitted signal is detected via a 4K low noise HEMT amplifier providing the cavity transmission coefficient $S_{21}$. The experimental setup is explained in detail in previous experiments\cite{Goryachev:2014aa,Goryachev:2014ab,goryachev2015discovery}. 

The cavity transmission $S_{21}$ is related to the coupling strength between the spin ensemble and a particular cavity mode, $g_\textrm{eff}$, which can be solved by fitting the cavity transmission signal $S_{21}$: 
\begin{equation}
S_{21}(f)=\frac{\kappa_{c}}{(\kappa_{c}+\frac{\gamma_{d}}{2})-i(f-f_{0})+\frac{g_\textrm{eff}^{2}}{\frac{\gamma_{s}}{2}-i(f-f_{s})}}
\label{S21}
\end{equation}
where $f$ is the input microwave frequency, $f_{0}$ is the frequency of the cavity mode, $f_{s}$ is the frequency of the spin transition, which is a function of the applied DC magnetic field, $\kappa_{c}$ is the coupling of probes to the cavity mode, $\gamma_{d}$ is the dielectric loss of the cavity mode, and $\gamma_{s}$ is spin loss. One may determine $g_\textrm{eff}$ by fitting the maximum of a transmission curve whilst scanning magnetic field and signal frequency before and after the LZ assisted thermalisation, or investigate the population dynamics by driving the system on the cavity resonance tuned on the $\ket{+7/2}$ to $\alpha \ket{-3/2} + \beta \ket{+5/2}$ level splitting in the cavity driven thermalisation approach. By fitting the dependence of the resonance frequencies of $S_{21}$ on the external field, one  effective coupling parameter is derived based on many measured data points covering all significant figures that are given in the results.

\subsection{Landau-Zener Assisted Thermalisation}

WGH$_{311}$ (9.45GHz) and WGH$_{211}$ (7.46GHz) are selected to couple to spin transition $\ket{-7/2}\rightarrow\ket{-5/2}$ (Probe 1) and $\ket{+7/2}\rightarrow\ket{+5/2}$ (Probe 2) at 26mT and 43mT respectively(see Fig.~\ref{figB}). Thus, Probes 1 and 2 give information on the occupancy of levels $\ket{-7/2}$ and $\ket{+7/2}$ respectively via estimated effective couplings $g_{\ket{\pm7/2}}$ as stated above. In order to observe the Landau-Zener assisted thermalisation, these quantities are measured before (increasing the magnetic field) and after (decreasing the field) tuning the spin ensemble through the avoided level crossing between $\ket{+7/2}$ and $\ket{-5/2}$ levels up to $100$~mT. The magnetic field tuning speed is set to $2$~mT/min, which is slow enough to ensure completeness of all transients and to avoid any heating effects. The probing field is either kept minimal (much less than the number of impurity ions) or switched on only for the fields when the spin ensemble is tuned on the cavity resonance, so as to avoid any confusion with pumping effects. As a result, we obtain four values of the coupling strengths: $g_{\ket{\pm7/2}}^\uparrow$ for the ensemble-cavity interaction before the LZ transition and $g_{\ket{\pm7/2}}^\downarrow$ after the LZ assisted relaxation. This allows us to estimate corresponding level occupancies before $N_{\ket{\pm7/2}}^\uparrow$ and after $N_{\ket{\pm7/2}}^\downarrow$ the event. Both numbers must add up to the total number of impurity spins $N$ as only these levels could be occupied at the temperature of $20$~mK:
\begin{equation}
\begin{array}{l}
N_{\ket{+7/2}}^\downarrow+N_{\ket{-7/2}}^\downarrow=N,\\
N_{\ket{+7/2}}^\uparrow+N_{\ket{-7/2}}^\uparrow=N.\\
\label{N}
\end{array}
\end{equation}
This relation is experimentally confirmed by the fact no interactions between the Gd$^{3+}$ ensemble and the microwave modes are observed in the cases when energy levels above $\ket{+7/2}$ play the role of the interaction lower energy state.

\begin{figure*}[htb]
\centering
\includegraphics[width=1.99\columnwidth]{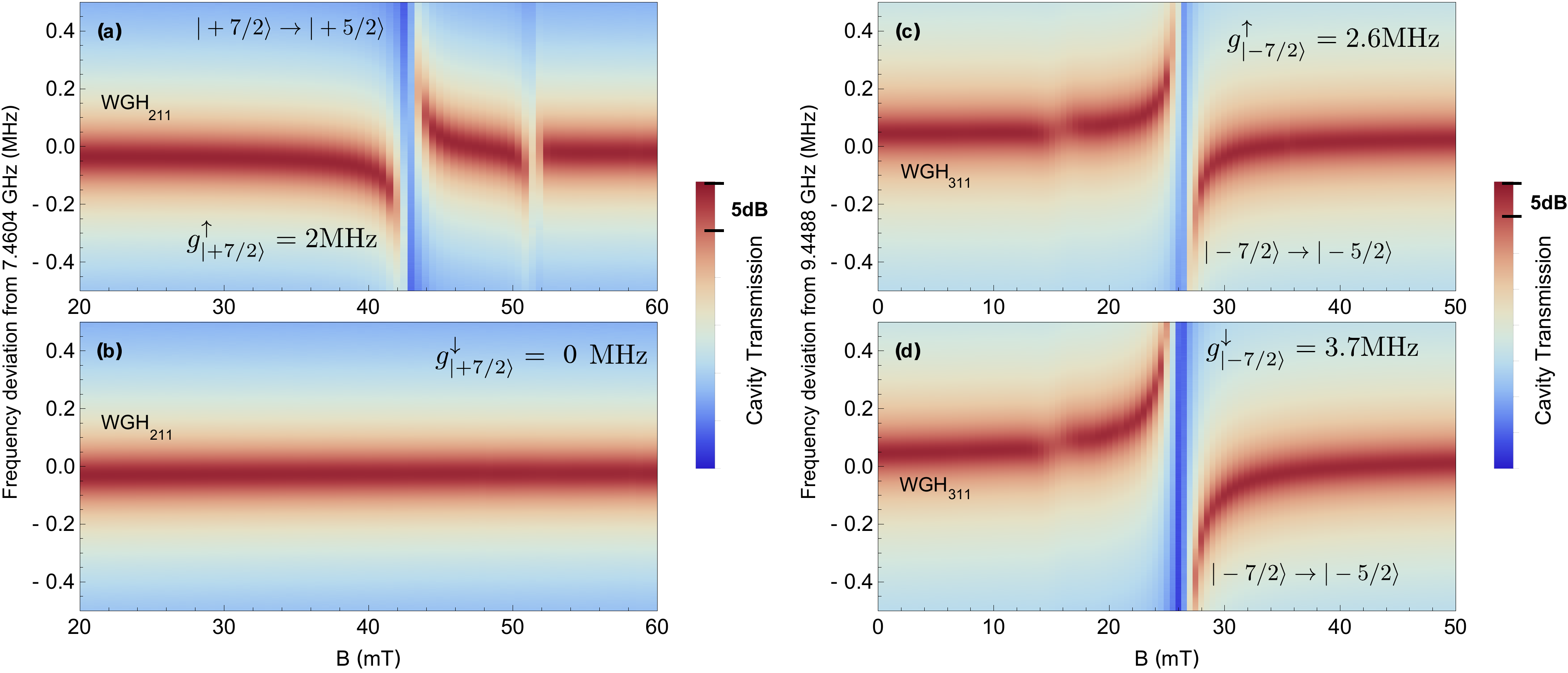}
\caption{Interaction between WGMs and Gd$^{3+}$: (a) WGH$_{211}$ and $\ket{+7/2}\rightarrow\ket{+5/2}$ spin transition before LZ transition. (b) WGH$_{211}$ and $\ket{+7/2}\rightarrow\ket{+5/2}$ after LZ transition, (c) WGH$_{311}$ and $\ket{-7/2}\rightarrow\ket{-5/2}$ before LZ transition, (d) WGH$_{311}$ and $\ket{-7/2}\rightarrow\ket{-5/2}$ after LZ transition.}
\label{hyster}
\end{figure*}

The result of the discussed measurement procedure in terms of the cavity transmission is shown in Fig.~\ref{hyster}. Plots (a) and (b) demonstrate complete leakage of the population from $\ket{+7/2}$ after tuning the system through the level interaction point. We see the corresponding ALC totally disappears and the corresponding coupling switches from $g_{\ket{+7/2}}^\uparrow=2$~MHz to $g_{\ket{+7/2}}^\downarrow=0$~MHz. On the other hand, density plots (c) and (d) confirm some increase in the ensemble-photon coupling from $g_{\ket{-7/2}}^\uparrow=2.6$~MHz to $g_{\ket{-7/2}}^\downarrow=3.7$~MHz (i.e. by a factor of $\sqrt{2}$), and thus the increase of the population on the ground state by a factor of 2. In terms of the occupancy, the following expected relations are fulfilled:
\begin{equation}
\begin{array}{l}
2N_{\ket{-7/2}}^\uparrow=N_{\ket{-7/2}}^\downarrow,\\
N_{\ket{+7/2}}^\downarrow=0.
\label{N2}
\end{array}
\end{equation}
It should be mentioned that the higher order mode WGH$_{311}$ is more confined inside the crystal, making coupling strength $g_{\ket{-7/2}}$ larger than $g_{\ket{+7/2}}$ even though $N_{\ket{+7/2}}=N_{\ket{-7/2}}$.

 
 The concentration of Gd$^{3+}$ ion impurities in the crystal can be estimated using coupling strength $g_{\ket{+7/2}}^{\downarrow}$:
\begin{equation}
n=\frac{N_{\ket{-7/2}}^\downarrow}{V}=\frac{g_{\ket{-7/2}}^{\downarrow2}}{g_{0}^{2} V} = 2.5\times10^{16}\textrm{cm}^{-3},
\end{equation}
where $V$ is the crystal volume and $g_{0}$ is estimated using Eq.~(\ref{g0}). This value is small enough to ignore direct spin-spin interaction between Gd$^{3+}$ ions thus ensuring the phase of the ensemble is paramagnetic\cite{Farr:2015aa}.

\begin{figure}[htb]
\centering
\includegraphics[width=0.87\columnwidth]{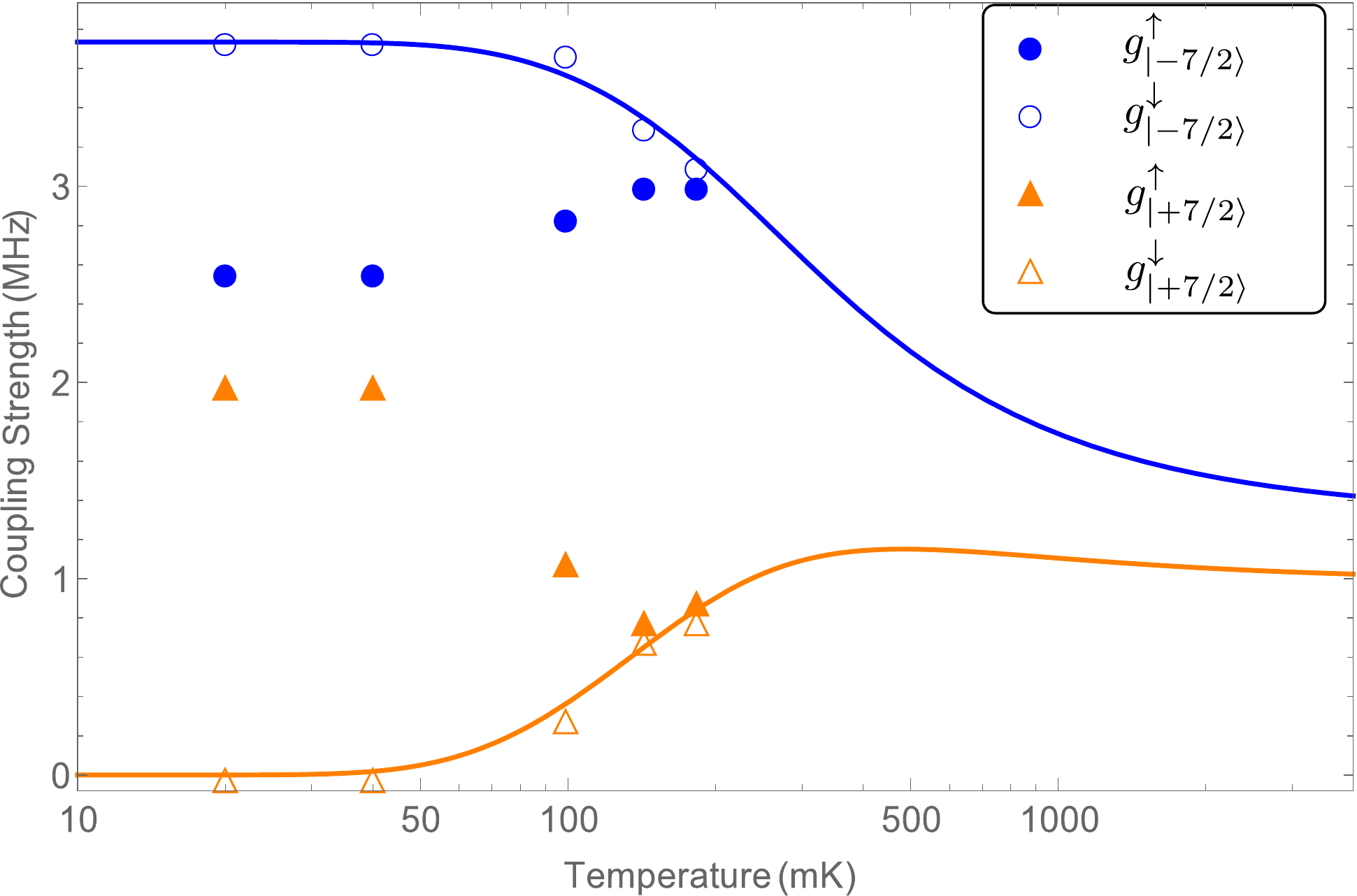}
\caption{Ensemble-cavity mode coupling strengths $g_{\ket{\pm7/2}}^\uparrow$ and $g_{\ket{\pm7/2}}^\downarrow$ as a function of temperature. The solid curves show predicted values for the system in thermal equilibrium state.}
\label{gtem}
\end{figure}

 $g_{\ket{\pm7/2}}^\uparrow$ and $g_{\ket{\pm7/2}}^\downarrow$ are further measured as a function of temperature  up to $200$~mK. The result is shown in Fig.~\ref{gtem} together with the calculated values for the case of thermal equilibrium (solid curves). The estimations are made assuming the spin energy level structure shown in Fig.~\ref{energylevel} is constant for all temperatures. These results confirm two predictions made above. First, measured coupling rates $g_{\ket{\pm7/2}}^\downarrow$ at all temperatures are consistent with those calculated in the case of thermal equilibrium, suggesting that the ion ensemble after the LZ transition is in thermal equilibrium state. Second, measured $g_{\ket{\pm7/2}}^\uparrow$ below $200$~mK are distinct from the thermal equilibrium state, suggesting that the thermalisation process between $\ket{\pm7/2}$ is negligible as described in the previous section. 
Moreover, when the temperature is below $50$~mK, these results correspond to the ensemble distribution shown in Eq.~\ref{N} . Further theoretical investigation of the dynamics for the spin system presented in this work may be performed using approaches used to study Landau-Zener transitions of a single TLS in a dissipative environment\cite{Nalbach:2009aa,Nalbach:2010aa}.


 
\subsection{Cavity Driven Thermalisation}
 
\begin{figure}
\centering
\includegraphics[width=0.87\columnwidth]{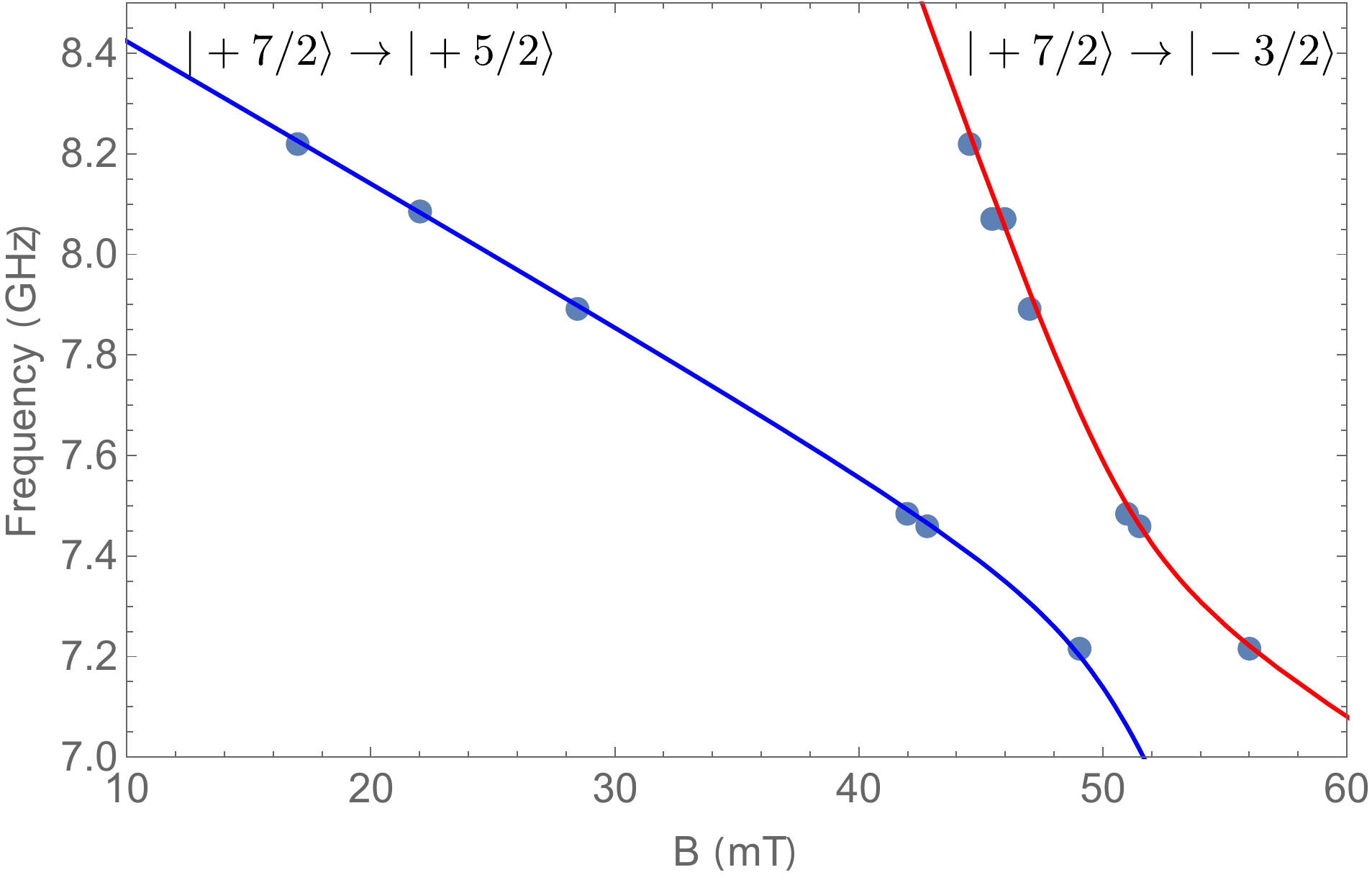}
\caption{Interaction of spin transitions from $\ket{+7/2}$ to the hybridised states $\ket{+5/2}$ and $\ket{-3/2}$. The solid curves are the calculated transition frequencies.}
\label{4thg}
\end{figure}

Electron spin resonance spectroscopy using WGMs\cite{Farr:2013aa} of the YVO$_4$ crystal at $20$~mK revealed the structure of the level interaction of Gd$^{3+}$ ions. The result is shown in Fig.~\ref{4thg} where both 
transitions $\ket{+7/2}\rightarrow\ket{+5/2}$ and $\ket{+7/2}\rightarrow\ket{-3/2}$ are observed. Five-photon transition $\ket{+7/2}\rightarrow\ket{-3/2}$ becomes visible due to the final state hybridisation $\ket{+5/2}$, allowing a single photon transition. The coupling strength between $\ket{+5/2}$ and $\ket{-3/2}$ is estimated to be $220$~MHz giving 5\% mixture of $\ket{-3/2}$ with $\ket{+5/2}$ at $43$~mT. The solid curves in Fig.~\ref{4thg} show the estimated spin transitions from the spin Hamiltonian with the $220$~MHz level coupling.

\begin{figure} 
\centering
\includegraphics[width=0.87\columnwidth]{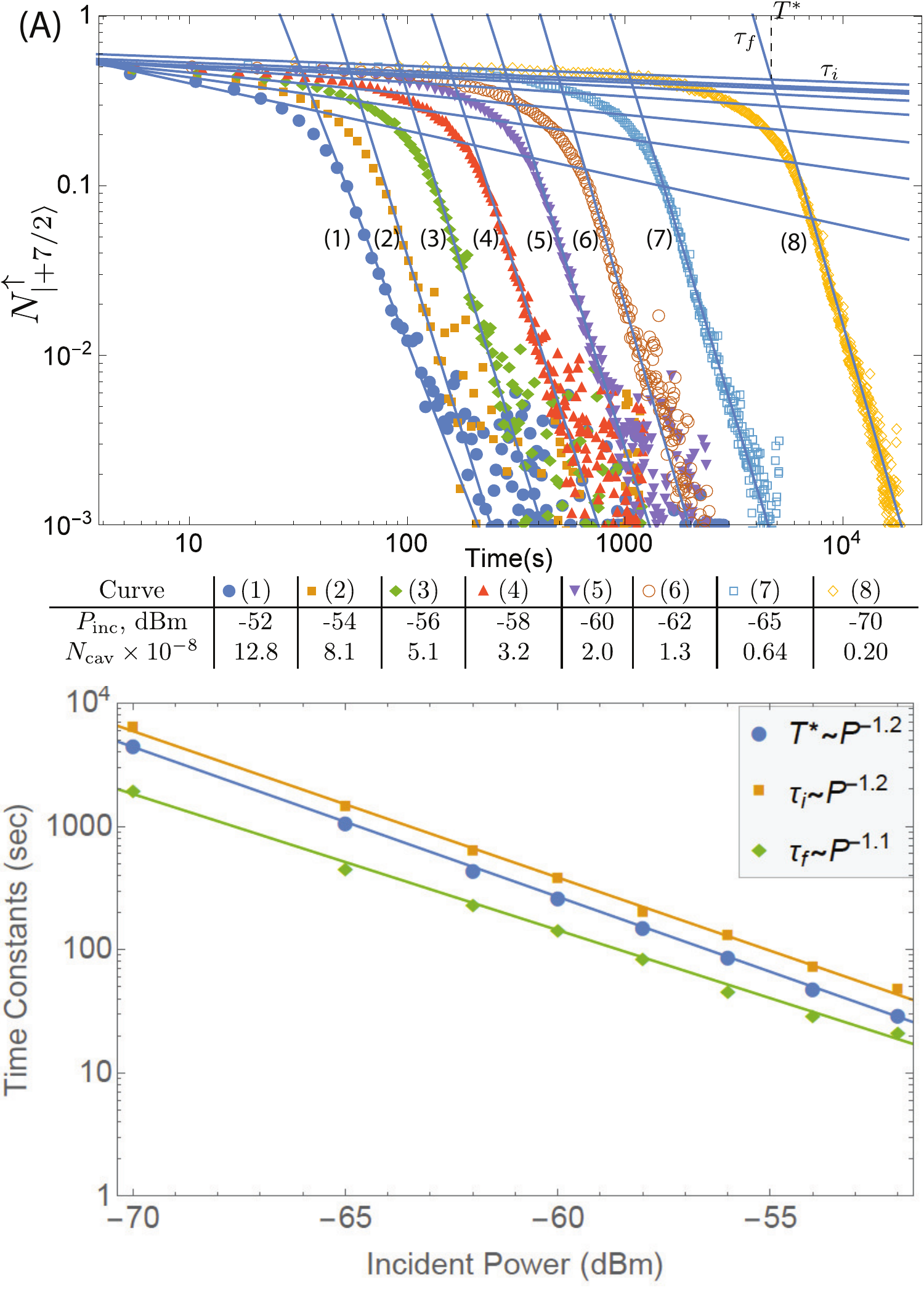}
\caption{(A) Decay of population of $\ket{+7/2}$ state for continuously pumped cavity for different values of the incident power (see Fig.~\ref{FigC}). (B) Dependence of the time constants of the decay processes on the incident power.}
\label{decay}
\end{figure}

As described in the previous section, the cavity driven thermalisation may be observed by tuning the hybridised spin transition onto the WGH$_{211}$ mode and pumping at the cavity resonance. The measured transmission in this case is found from (\ref{S21}) by putting $f=f_{0}=f_{s}$ and $g_\textrm{eff}=g_{0}\sqrt{N_{\ket{+ 7/2}}^\uparrow}$:
{\begin{equation} 
S_{21}(f_{0})=\frac{\kappa_{c}}{\kappa_{c}+\frac{\gamma_{d}}{2}+\frac{2 g_{0}^{2}}{\gamma_{s}} N_{\ket{+ 7/2}}^\uparrow},
\label{S210}
\end{equation}}
where $\kappa_{c}$, $\gamma_{d}$, $g_{0}$ and $\gamma_{s}$ are constants, giving a dependence of the observed quantity on the occupation of the first excited level.

Time dependence of transmission in Eq. (\ref{S210}) is measured at $43$~mT for different values of the incident power $P_\textrm{inc}$ giving different number of stored photons $N_\textrm{cav}$. The measured transmission is converted into population $N_{\ket{+ 7/2}}^\uparrow$ and shown in Fig.~\ref{decay}. The result suggests that the population of the ${\ket{+ 7/2}}$ level decays to zero thus approaching thermal equilibrium as described in the previous section. Each observed decay may be separated into two exponential processes giving three time parameters: time constant of the initial decay $\tau_i$, time constant of the finial decay $\tau_f$, transition time from one into the other $T^\star$. It is observed that $\tau_i$ increases with decreasing power

To explain these results, the energy level diagram may be approximately reduced to the well known $\Lambda$-scheme with levels $\ket{1} = \ket{+7/2}$, $\ket{2} =\alpha\ket{-3/2}+\beta\ket{+5/2}$ and $\ket{3} = \ket{-7/2}$ (see Fig.~\ref{FigC}). With $\mathbf{N}$ being the vector of populations of corresponding levels, the system dynamics might be described by the following set of ordinary differential equations:
\begin{equation}
\dot{\mathbf{N}} = \left(\begin{array}{ccc}
  -W(N_\textrm{cav}) & \Gamma_{21} & 0 \\
  W(N_\textrm{cav}) &  -\Gamma_{21} - \Gamma_{23} & 0  \\
  0 & \Gamma_{23} & 0 \end{array}  \right)
  {\mathbf{N}} 
\label{N2}
\end{equation}
where $W(N_\textrm{cav})$, $\Gamma_{21}$ and $\Gamma_{23}$ are time constants associated with the cavity pump from $\ket{1}$ to $\ket{2}$ and decay from $\ket{2}$ to $\ket{3}$ and $\ket{1}$. In the limit of strong pumping ($W\gg \Gamma_{21}$ and $W\gg\Gamma_{23}$), two eigenvalues that can be observed through the population of $\ket{1}$ are estimated as $-\frac{1}{2}(\Gamma_{21}+\Gamma_{23})$ and $W$. These values are estimations of the attenuation rates corresponding to $\tau_f$ and $\tau_i$ time constants correspondingly observed in the experiment.
This result predicts a small power dependent time constant (slow decay) and a large almost power independent time constant (fast decay) that matches the experimental observations in Fig.~\ref{decay}.




\section{Discussion}

{The unconventional level structure of Gd$^{3+}$ ions in YVO$_{4}$ crystal with the two lowest energy levels split by a seven photon transition provides a qubit system with virtually {\it  unlimited relaxation time}. Indeed, truncating the structure between $\ket{+7/2}$ and $\ket{-5/2}$ levels, one comes up with a qubit whose most general state may be written as
\begin{equation} 
\ket{\phi} = \alpha\ket{-7/2}+\beta\ket{+7/2},
\label{qubit}
\end{equation}
where $\alpha,\beta\in\mathbb{C}$ and $\ket{-7/2}$ being the ground state. Moreover, at readily accessible temperature of around $20$mK and some external magnetic field, the energy splittings are such that only the ground state is populated at thermal equilibrium. And, finally, the energy splitting of a such qubit is externally controllable by the magnetic field. Such properties could be extremely fruitful for implementation of quantum memories at microwave frequencies that nowadays suffer from relatively low coherence times. Although, despite the fact that a Gd$^{3+}$:YVO$_{4}$ memory provides unlimited relaxation times, its coherence times should be still experimentally determined. Additionally, such a system cannot be efficiently addressed directly as it would require some complicated protocols involving higher energy levels and, probably, manipulations with the DC external field. 

Also, Landau-Zener-St{\"u}ckelberg interferometry based on periodic passage of an avoided level crossing responsible for the LZ transition, has been found to be a useful tool for studying properties of atomic and superconducting qubit systems\cite{Shevchenko:2010aa}, including perfect population transfer, implementation quantum gates for quantum-control and quantum-computing\cite{Teranishi:1998aa,Gaitan:2003aa,Nagaya:2007aa}.

Another potential area of application for the considered system is microwave-to-optical quantum converters\cite{OBrien:2014aa,Williamson:2014aa} required for coherent information exchange between local quantum processors. In addition to the advantages of Gd$^{3+}$:YVO$_{4}$ described above, the very existence of large zero field splitting is advantageous for such applications, since such hybrid system would not require large external magnetic field that can negatively influence superconducting parts. Moreover, excellent optical properties of "vanadate lasers" (rare earth ions doped YVO$_4$ and GdVO$_4$) are widely exploited nowadays mostly in the form of Nd$^{3+}$:YVO$_4$\cite{Connor,Pavel:2008aa}.  Although, optical transitions of Gd$^{3+}$ ions in vanadates have to be studied separately. 

The quantum level structure of Gd$^{3+}$ in YVO$_{4}$ at zero field is suitable for physical realisation of a continuous wave maser using the approach similar to that of Fe$^{3+}$ ions in sapphire\cite{Creedon:2010aa}. Indeed, at zero field the lower six levels are degenerate in three pairs and form the three level system: $\ket{\pm7/2}$, $\ket{\pm5/2}$, $\ket{\pm3/2}$. By externally pumping the upper level $\ket{\pm3/2}$ from the ground state $\ket{\pm7/2}$, one can induce the maser signal associated with one of the decaying process. Such maser can be built based on the employed WGM approach by designing the cavity with a resonant mode corresponding to one of the decay processes. A maser system can be also realised at nonzero external magnetic field using, for example, the scheme similar to that in Fig.~\ref{FigC} and, thus, exploiting the coupling between energy levels. In this case, the resonator will also require to have a mode at a desired frequency. In both cases, concentration of Gadolinium ions in the resonator mast be increased. 

Another possible application of the Gd$^{3+}$:YVO$_{4}$ system exhibiting the spin angular momentum limitations on some transitions is in microwave detectors that potentially can reach quantum sensitivities. For such device that is highly required for many quantum optics experiments at microwave frequencies, a $\Lambda$-scheme has been proposed\cite{qradar,Inomata:2016aa}. Such scheme requires long lived ground and the first excited states together with short lived second excited state. These requirements are fulfilled within the cavity assisted thermalisation scheme depicted in Fig.~\ref{FigC}. Here the long lived first excited state $\ket{+7/2}$ is the so-called initial state. By absorbing an external photon the system transients into the exited state $\alpha \ket{-3/2} + \beta \ket{+5/2}$ from which it decays into the lowest energy state $\ket{-7/2}$. By monitoring population of this level, or equally $\ket{+7/2}$, with some external macroscpoic parameter depending on the population of this levels, one can count the number of incoming photons. In principle, the cavity assisted thermalisation experiment demonstrated in this work may be re-interpreted in this way. Here the role of the external macroscopic parameters is played by the cavity transmission and ensemble spin-photon coupling. Although, the demonstrated realisation is not in the desired single-photon regime. In order to achieve such regime one would require to reduce the cavity linewidth or use some other macroscopic parameters. 
}

In summary, using Gd$^{3+}$ doped YVO$_{4}$ crystal, we demonstrate peculiarities of QED based on $7/2$ spin system with the ground and the first excited states separated by a seven photon transition. In particular, it is demonstrated that by sweeping external magnetic field from zero, one may generate a non-equilibrium state of the spin ensemble with extremely long life times due to large number of photons required for the transition between the ground and the first excited states. Then, it is shown that the equilibrium state may be achieved through the Landau Zener transition or external cavity drive, being possible due to coupling between excited energy levels. The experiments signify the importance of a particular realisation of a spin ensemble for Quantum Electrodynamics.

This work was supported by the Australian Research Council grant number CE110001013.

\hspace{15pt}

\section*{References}


%

\end{document}